\documentclass[reprint,showpacs,floats,aps,unsortedaddress,superscriptaddress]{revtex4-1}
\usepackage{graphicx}
\usepackage{color}
\usepackage{hyperref}
\hypersetup{colorlinks=true, allcolors=blue, breaklinks=false}

\newcommand{\insertFig}[4]{
  \includegraphics[width=#1]{#2} 
  \caption{\label{fig:#3}{#4}}
}

\newcommand{\bYAB}{$\beta$-YbAlB$_4$}
\newcommand{\aYAB}{$\alpha$-YbAlB$_4$}
\newcommand{\Zt}{\widetilde{Z}}
\newcommand{\ket}[1]{$\left|#1\right>$}
\newcommand{\Tcoh}{T_{\mathrm{coh}}}
\newcommand{\TK}{T_{\mathrm{K}}}

\begin{document}

\title{Kondo hybridization and quantum criticality in $\beta$-YbAlB$_4$ by
laser-ARPES}

\author{C\'{e}dric~Bareille}
\email{bareille@issp.u-tokyo.ac.jp}
\affiliation{ISSP, University of Tokyo, Kashiwa 277-8581, Japan}

\author{Shintaro~Suzuki}
\affiliation{ISSP, University of Tokyo, Kashiwa 277-8581, Japan}

\author{Mitsuhiro~Nakayama}
\affiliation{ISSP, University of Tokyo, Kashiwa 277-8581, Japan}

\author{Kenta~Kuroda}
\affiliation{ISSP, University of Tokyo, Kashiwa 277-8581, Japan}

\author{Andriy~H.~Nevidomskyy}
\affiliation{Department of Physics and Astronomy \& Center for
Quantum Materials, Rice University, Houston, Texas 77005, USA}

\author{Yosuke~Matsumoto}
\affiliation{Max Planck Institute for Solid State Research,
Heisenbergstrasse 1, Stuttgart 70569, Germany}

\author{Satoru~Nakatsuji}
\affiliation{ISSP, University of Tokyo, Kashiwa 277-8581, Japan}
\affiliation{CREST, Japan Science and Technology Agency (JST),
4-1-8 Honcho Kawaguchi, Saitama 332-0012, Japan}

\author{Takeshi~Kondo}
\affiliation{ISSP, University of Tokyo, Kashiwa 277-8581, Japan}

\author{Shik~Shin}
\affiliation{ISSP, University of Tokyo, Kashiwa 277-8581, Japan}

\begin{abstract}
We report an angle-resolved photoemission (ARPES) study of \bYAB, which
is known to harbor unconventional quantum criticality (QC) without any tuning.
We directly observe a quasiparticle peak (QP),
emerging from hybridization, characterized by a binding energy and an onset of
coherence both at about 4 meV. This value conforms with a previously
observed reduced Kondo scale at about 40~K.
Consistency with an earlier study of carriers in \bYAB~via the Hall effect
strongly suggests that this QP is responsible for the QC in \bYAB.
A comparison with the sister polymorph \aYAB, which is not quantum critical
at ambient pressure, further supports this result. 
Indeed, within the limitation of our instrumental resolution, our ARPES
measurements do not show tangible sign of hybridization in this locally
isomorphic system, while the conduction band we 
observe is essentially the same as in \bYAB. 
We therefore claim that we identified by ARPES the carriers responsible
for the QC in \bYAB.
The observed dispersion and the underlying hybridization of this QP are
discussed in the context of existing theoretical models.
\end{abstract}

 \maketitle 

\section{Introduction}
The formation of a non-magnetic Kondo-singlet ground-state commonly provides
a description of heavy Fermi liquid (FL) in intermetallics rare-earth compounds. 
It emerges below the Kondo temperature $\TK$, from the screening of the local
moments of $f$ orbitals by conduction electrons, which accompanies the onset
of coherence of a quasiparticle (QP) peak binding at $k_{\mathrm{B}}\TK$:
the Kondo resonance peak~\cite{Hewson_Kondo2HF}. 
When external parameters are accurately tuned, numerous heavy fermion systems
deviate from this normal FL. 
In the conventional scenario, spin-density-waves instabilities induce such
non-FL behavior by driving the system to the critical regime of
a magnetic quantum transition~\cite{Lohneysen_RMP2007}. 

In \bYAB, quantum criticality (QC) has been observed below about 4~K
at zero magnetic field and ambient pressure~\cite{Matsumoto_Science2011},
away from any magnetic order~\cite{Tomita_Science2015}.
It, therefor, can not be cast in this usual picture of magnetic instability. 
Moreover it shows
an unexpected ($T/B$)-scaling~\cite{Matsumoto_Science2011, Matsumoto_JPSJ2015}
and unusual critical exponents~\cite{Nakatsuji_NatPhys2008},
which challenge theoretical descriptions. 
Diverse works~\cite{Misawa_JPSJ2009, Ramires_PRL2012, 
Pixley_PRL2012, Hackl_PRB2011, Watanabe_JPSJ2014, Shaginyan_PRB2016}
attempt to model this unconventional QC, without any possible consensus. 
Although the FL behavior is recovered with a moderate magnetic field, 
its description and formation in term of Kondo screening
are still poorly understood. 
Indeed, despite a large Kondo scale $\TK \approx 250$~K, incoherent skew
scattering is observed down to $\Tcoh \approx 40$~K~\cite{OFarrell_PRL2012},
and magnetic moments are still present
as low as $T^* \approx 8$~K~\cite{Matsumoto_JPSJ2015}. 
Kondo lattice behavior finally sets in only below $T^*$. 
Advanced analysis of Hall effect measurements shed some light
by the identification of two different components of carriers
with distinct Kondo scales~\cite{OFarrell_PRL2012},
which we call carriers $a$ and carriers $b$ for convenience. 
Carriers $a$ govern the longitudinal transport with $\TK$ as the Kondo scale
and a relatively high carrier density $n_a^\mathrm{Hall}$, 
while carriers $b$ have a low carrier density $n_b^\mathrm{Hall}$
with electronlike character
and gains coherence below a second Kondo scale at $\Tcoh$. 
The conclusion from Hall effect measurements further identifies carriers $b$
as responsible for both Kondo lattice and QC behaviors
below $T^*$~\cite{OFarrell_PRL2012}. 
Nevertheless, the details about the emergence of these carriers 
and about the underlying Kondo physics are still lacking. 

In the present article, we take advantage of the apparent location of the
QC at zero magnetic field, without tuning of any control
parameter~\cite{Matsumoto_Science2011}, to directly probe QP in
\bYAB~by high-resolution laser-ARPES. 
In order to highlight any relation between microscopic observations and the QC,
we also perform ARPES measurements on the sister compound \aYAB, whose structure
is sketched alongside that of \bYAB~in Fig.~\ref{fig:kzDep}(a). 
This locally isomorphic compound exhibits very similar behaviors and
temperature scales, however unlike \bYAB, it does not display QC,
developing instead a heavy Fermi liquid~\cite{Matsumoto_PRB2011}. 
Moreover, it has been shown that the magnetization in \bYAB~can be separated
into two components: the QC component and a heavy Fermi-liquid component similar
to that in \aYAB~\cite{Matsumoto_JPSJ2015}. 
Thus, \aYAB~is a perfect control system to study the unconventional criticality
in \bYAB. 

Our measurements reveal the hybridization between the crystal-field split heavy
Yb $f$-electron bands and a light conduction band in \bYAB. 
The resulting electronlike QP has a binding energy and
an onset of coherence at about about 4~meV, consistent with carriers $b$
highlighted by Hall measurements as responsible for the QC.
By contrast, we find almost no or very weak hybridization in the sister
$\alpha$-phase, while the dispersion of the conduction band is very similar
between the two phases. 
Our results strongly suggest the importance of the Kondo hybridization
for the microscopic mechanism behind the QC in \bYAB. 

\section{Methods}
High-quality single crystals were grown by using the Al-flux method as described
in the literature~\cite{Macaluso_CoM2007}. The excess Al flux was etched
by a water solution of sodium hydroxide.
All ARPES measurements were performed with a SES R4000 analyzer, in a chamber
shielded from magnetic field by mu-metal. Clean surfaces were obtained by
cleaving crystals along the $(a,b)$ plane in ultra-high vacuum
(below $5\times10^{-11}$~Torr). 
Measurements in the (100)-plane were carried out at the 1-cubed end-station at
Helmholtz-Zentrum Berlin, by tuning the photon
energy from 24~eV to 120~eV, thanks to the synchrotron radiation of the BESSY-II
facility. The temperature was about 1~K, and the overall resolution was about
10~meV. 
Measurements in the (001) plane for both $\beta$ and $\alpha$ phases, were
realized with a 6.994~eV laser at the ISSP, at a temperature of 5~K, with an
overall resolution of about 3~meV. 
Data from \bYAB~were measured with circular polarized light, while data from
\aYAB~were measured with linear horizontal polarized light. Nevertheless, as
Fig. S1 of the Supplemental Material shows~\cite{SuppMat}, 
measurements with linear horizontal
polarized light give similar results to those with circular polarized light. 
Thus, this difference does not affect our observations. 
Treatments by 2D curvature of some spectra of this work are performed
following the method described in Ref.~\onlinecite{CurvaturePeng}.

\section{Results}
The \bYAB~crystal has an orthorhombic structure
\cite{Macaluso_CoM2007}; Fig.~\ref{fig:kzDep}(b) schematizes the
corresponding Brillouin-zone (BZ).
Samples were cleaved along the $(a,b)$ plane to perform photoemission. 
Fig.~\ref{fig:kzDep}(c) shows the edges of the BZ overlaid on a mapping in the
(100) plane, 50~meV below the Fermi level, while the dispersion along the normal
emission, $\Gamma$-$Z$ line, is displayed on Fig.~\ref{fig:kzDep}(d).
Conversion into momentum space was made in the frame of the nearly-free-electron
final-state approximation~\cite{Damascelli_2004}.
The inner potential was determined to be $V_0 \approx 22$~eV thanks to the
point-group symmetry and periodicity, with, as a main indication, a hole-like
band centered on the $Z$ point, depicted by the gray parabola in
panel~(d) and the gray ellipse in panel~(c).
Also visible in panel~(d) is a nearly flat band bounded at about 100~meV below
$E_\mathrm{F}$ -- it is a part of the Yb$^{2+}$ $|J = 7/2\rangle$ multiplet,
which has been previously observed by X-ray photoemission spectroscopy
\cite{Okawa_PRL2010}. 
Complete investigation and discussion over the full three-dimensional BZ will be
part of a future work. Here, we will  focus instead on the results from
laser-ARPES, which, while being more restricted in $\mathbf{k}$ space, offers
very high resolution and intensity. 

Given the laser photon energy $h\nu = 6.994$~eV, the available BZ
cut is depicted by the lower purple circle in Fig.~\ref{fig:kzDep}(c), which
crosses the $\Gamma$-$Z$ line less than 0.2~\AA$^{-1}$ below the
Z point, named $\Zt$ for convenience. 
Measurements with photon energy $h\nu = 54$~eV reach an equivalent position
along the $\Gamma$-$Z$ line [see the higher purple circle in
Fig.~\ref{fig:kzDep}(c)].
The dispersive hole-like band with the top bound at about $-45$~meV,
as well as the flat band at about $-90$~meV, were observed using
the light source both with photon energy $h\nu = 54$~eV
[see panels (c) and (d)] as well as with the laser in panel (e), 
thus providing a calibration of our laser-ARPES.
In what follows, we focus on the bands above $E>-40$~meV, which are observable
thanks to laser-ARPES, and we compare them to our measurements on \aYAB. 

We investigate the vicinity of $E_\mathrm{F}$ with the stack of energy
distribution curves (EDCs), Fig.~\ref{fig:disp5K}(a), extracted along 
$k_\mathrm{y}$, i.e., along the analyzer slit, with $k_\mathrm{x} = 0$
[full gray line $A$ on Fig.~\ref{fig:disp5K}(b)]. 
We observe three dispersions; they can be modeled by a light electronlike band
with $m^\ast\approx 3~m_e$ and the band bottom of $\approx 20$~meV below
$E_\mathrm{F}$, which hybridizes and forms anticrossings with two flat bands,
a first one around $-4$~meV and a second one around $-13$~meV
[thin dashed lines in Figs.~\ref{fig:disp5K}(c) and \ref{fig:disp5K}(d)]. 
These two flat bands, separated by 9~meV are consistent with the crystal
electric field of about 80~K, extracted from fitting the magnetic susceptibility
\cite{Nevidomskyy_PRL2009}. From this fitting and local symmetry considerations
detailed in Ref.~\onlinecite{Nevidomskyy_PRL2009}, 
the two flat bands can then be
identified as the ground-state $J_z=\pm5/2$ (\ket{\pm5/2}), at $-4$~meV, and the
first crystal-field excited state, dominated by $J_z=\pm3/2$ (\ket{\pm3/2}), at
$-13$~meV, of the J = 7/2 Yb multiplet. 
In Fig.~\ref{fig:disp5K}(c), solid lines are the results of fitting to the
hybridization model with a constant hybridization function  $V \approx 4$~meV. 
The first anti-crossing at $-5$~meV is nicely fitted, and forms an electronlike
QP band [blue line in Fig.~\ref{fig:disp5K}(c)] bound at about 4~meV
below $E_\mathrm{F}$. 
This energy scale, one order of magnitude smaller than $\TK \approx 250$~K,
is consistent with the expectation from previous experimental
observations~\cite{OFarrell_PRL2012,Matsumoto_PRB2011} of a reduced Kondo scale
at $\Tcoh \approx 40$~K~($\approx 3.4$~meV). 

The temperature dependence of EDCs in Fig.~\ref{fig:Tdep} confirms
the relation between the QP peak and $\Tcoh$. 
Panels (a) and (b) shows EDCs from 5~K to 60~K
at the inner ($k_\mathrm{y} = 0$~\AA$^{-1}$) 
and outer ($k_\mathrm{y} = 0.17$~\AA$^{-1}$) sides of the anti-crossing
respectively. The same EDCs divided by the Fermi-Dirac step are displayed
at the bottom of these two panels. 
For both, the QP peak slowly disappears as the temperature goes up. It is
clearer on the outer EDCs where the peak intensity suddenly drops
from 30~K to 40~K. 
The peak is totally restored by cooling back down (light blue EDC), showing that
our observations are not due to surface degradation. 
For both inner and outer peaks, the intensity integrated along the shaded area
[see panels (a) and (b), respectively] of EDCs divided by the Fermi-Dirac step
is displayed against the temperature in panel (c),
relatively to the value at 60~K. 
Both peaks slowly drop when temperature crosses $\Tcoh$. 
It is worth to emphasize that, to our knowledge, it is the first direct and
fully resolved observation by ARPES of a Kondo-like hybridization
with unified binding energy, onset of coherence, and Kondo scale. 
While a similar observation has been reported on YbRh$_2$Si$_2$,
both the QP peak and the hybridization gap
were not fully resolved~\cite{Mo_PRB2012}. 

Nevertheless, here, the second anti-crossing at $E\approx -15$~meV is badly
reproduced by the constant hybridization model
[pink line in Fig.~\ref{fig:disp5K}(c)].
By taking $\mathbf{k}$-dependent hybridization proposed theoretically
\cite{Nevidomskyy_PRL2009,Ramires_PRL2012}, we obtain a more satisfying fit,
as shown by the pink line in Fig.~\ref{fig:disp5K}(d).
The significance of $\mathbf{k}$-dependent hybridization is developed
in the discussion section. 
The fitting parameters of both models are detailed in the Supplemental
Material~\cite{SuppMat}. 
Finally, the hybridization with the \ket{\pm5/2} level at about $-4$~meV results
in the small electronlike QP peak [blue line in panels (c) and (d)
of Fig.~\ref{fig:disp5K}], which crosses the Fermi level at
$k^{\beta}_{\mathrm{F1}} \approx 0.09~$\AA$^{-1}$ and has
a mass of about $3~m_e$ at $E_\mathrm{F}$. It forms quite an isotropic
Fermi sheet, as can be seen via the in-plane mapping at $E_\mathrm{F}$ in
Fig.~\ref{fig:disp5K}(b)(blue circle). 
The size, mass, and isotropy of this small pocket are compatible with the
observations by quantum oscillations on \bYAB~\cite{OFarrell_PRL2009}. 

It should be noted that, along the $\left<010\right>$ direction, the
\ket{\pm5/2} level crosses $E_\mathrm{F}$ at about
$k^{\beta}_{\mathrm{F2}} \approx 0.25$~\AA$^{-1}$. 
This second Fermi-momentum is exhibited by red arrows on the Fermi-mapping,
Fig.~\ref{fig:disp5K}(b), as well as on the cut along the $\left<010\right>$
direction, Fig.~\ref{fig:disp5K}(e). 
Blue, red and pink full lines are guides to the eyes for the same dispersions as
Fig.~\ref{fig:disp5K}(d). 
The observation of a second Fermi-momentum implies an overlapping of the blue
and red hybridized bands, on the Fig.~\ref{fig:disp5K}(e). 
In order to account for it, we need to introduce a small
dispersion to the \ket{\pm5/2} level in our model, corresponding to a mass
$m_f^* \approx 70~m_e$ in the $\left<010\right>$ direction. 
This very large mass makes it difficult to be observed by quantum oscillations.

Carrier density can be estimated at the $T=0$ limit assuming spherical Fermi
sheets by $n=k_\mathrm{F}^3/3\pi^2$~~\cite{AshcroftMermin}.
Before hybridization, the little conduction band has a density of
$n_c^\mathrm{ARPES} \approx 7.4\times10^{25}$~m$^{-3}$. 
It is of the same order as the value estimated
from Hall measurements~\cite{OFarrell_PRL2012} of carriers $b$ at $T = 140$~K
(i.e., above the integration of the $f$ electron to the Fermi surface):
$n_b^\mathrm{hall}(T = 140$~K$)\approx 16\times10^{25}$~m$^{-3}$. 
Below $\Tcoh$, the carrier density could not be accurately extracted
from the Hall measurements.
Yet, estimation of the density of the two Fermi sheets from our measurements
(i.e., including $f$ electrons), 
$n^\mathrm{ARPES} \approx 55\times10^{25}$~m$^{-3}$, is still one order
of magnitude smaller than the density of carriers $a$
identified as the main contribution
to the transport by Hall effect measurements, estimated at 50~K
to $n_a^\mathrm{hall}(T = 50$~K$)\approx 416\times10^{25}$~m$^{-3}$. 
Therefore, carriers of \bYAB~observed here, 
thanks to the high resolution of laser-ARPES, match carriers $b$ identified
as being responsible for both Kondo lattice and QC behaviors
below $T^*$ by Hall effect measurements.
Specifically, they are of electronlike character, gain coherence below $\Tcoh$,
and have low carrier density.
From these evidences, we argue that the observed Fermi sheets and the underlying
Kondo hybridization are closely related to both Kondo lattice
and QC behaviors below $T^*$. 

To verify our claim,
we now look at the electronic structure of the sister compound \aYAB~in the
vicinity of $E_\mathrm{F}$. 
Fig.~\ref{fig:aYAB}(a) displays the laser-ARPES curvature spectra along
the $\left<100\right>$ direction, where mainly two features are visible:
an almost flat band and a light electronlike band dispersing all the way up to
$E_\mathrm{F}$. 
The former is revealed by fitting the peaks in the EDCs (open circles) at about
$-13$~meV below $E_\mathrm{F}$. 
The dispersive electronlike band is better seen by fitting the peaks from the
momentum distribution curves (MDCs) [full circles in Fig.~\ref{fig:aYAB}(a)
and~\ref{fig:aYAB}(b)] down to $-12$~meV; 
the two peaks are difficult to distinguish at higher binding energies. 
The peaks are directly visible on the stack of MDCs, Fig.~\ref{fig:aYAB}(b). 
This light dispersion crosses $E_\mathrm{F}$ at
$k^{\alpha}_{\mathrm{F}} \approx 0.13$~\AA$^{-1}$.

In Fig.~\ref{fig:bVSa}(a), we compare the two aforementioned bands with the
ones observed in \bYAB. 
Intriguingly, they are identical to the dispersive electronlike band and the
flat \ket{\pm3/2} band derived from the $\mathbf{k}$-dependent hybridization
model used for \bYAB, copied onto panel~(a) with dashed lines. 
On the other hand, the \ket{\pm5/2} level is seemingly not observed in \aYAB.
A peak is visible at $k=0$, however it is very limited in momentum space and may
correspond to a different conduction band. 
We directly compare the EDCs from $\beta$- and \aYAB~in
Fig.~\ref{fig:bVSa}(b), in red and green lines, respectively. 
Red open arrows show the \ket{\pm5/2} level as a peak on the \bYAB~EDCs.
On \aYAB,~only shoulders are visible. When divided by the Fermi-Dirac step,
these shoulders disappear, and no peak is left, thus indicating the absence of
the \ket{\pm5/2} level in our data on \aYAB.
On the other hand, one would expect the crystal field splitting to be very
similar between these two locally isomorphic phases, and the absence of the
\ket{\pm5/2} level in our data could be a result of a bad sample surface
quality, which may especially affect the conclusions so close to $E_\mathrm{F}$.
What can be said with certainty, however, is that we do not observe any sign of
hybridization between either of the flat Yb $4f$ levels and the conduction
electron bands in \aYAB, within our instrumental resolution of $\sim 3$ meV.
Indeed, the light electronlike band disperses straight through the $-4$ meV
energy to cut through $E_\mathrm{F}$, while no gap is visible at about $-13$~meV
where one would expect the second anti-crossing with the Yb \ket{\pm3/2} level. 
This is evidenced by the EDCs in Fig.~\ref{fig:bVSa}(b): on \aYAB,
the \ket{\pm3/2} level (full green arrows) is at the same position on both
$k = 0$ and $k = \pm0.15$~\AA$^{-1}$, in clear contrast with EDCs on \bYAB.
We conclude that the manifestation of the Kondo hybridization
in the one-particle spectral function (see Ref.~\onlinecite{Damascelli_2004})
is much less pronounced in $\alpha$-YbAlB$_4$ compared to the $\beta$-phase,
although within the experimental resolution of our laser-ARPES measurements,
we cannot exclude manifestations of the hybridization smaller
than about $\sim~3$ meV in the $\alpha$-phase.

\section{Discussion}
The absence of tangible sign of Kondo hybridization in our ARPES data on \aYAB,
in contrast to \bYAB, further supports the idea that the hybridization
and emerging QP peak play a crucial role in the
unconventional quantum criticality observed in the latter compound.
Moreover, our analysis shows that the heavy dispersions of \bYAB~are best fitted
with a $\mathbf{k}$-dependent hybridization function, as put forward by the
theory of the critical nodal metal~\cite{Ramires_PRL2012}. 
The key feature of this latter model is that the Kondo hybridization vanishes
at the $\Gamma$ point quadratically in the in-plane momentum ($k_x$, $k_y$),
stemming from the $|\pm5/2\rangle$ nature of the $f$-level.
It implies the formation of a nodal line along the $c$ axis, which causes the non-FL behavior in $\beta$-YbAlB$_4$~\cite{Ramires_PRL2012}.
It is important to note however that for this model to explain the observed
$T/B$ scaling in \bYAB, the renormalized position $\tilde{E}_f$ of the ground
state \ket{\pm5/2} doublet must be pinned at the Fermi level. 
It is therefore difficult to unequivocally corroborate the theoretical model,
in its original form, based on our ARPES data.
It follows from our ARPES analysis that this doublet actually crosses the Fermi
level at the wavevector  $k^{\beta}_{\mathrm{F2}} \approx 0.25$~\AA$^{-1}$ and
has an intrinsic bandwidth, characterized by a heavy quasi-particle mass
$m_f^\ast \sim 70~m_e$. Clearly, this situation is more complex than in the
nodal metal theory of Ref.~\onlinecite{Ramires_PRL2012}, where the \ket{\pm5/2}
doublet was assumed to be perfectly flat, in other words the Yb-Yb electron
hopping was neglected. 
In this situation, it could be possible for a portion of the nodal line to be
bound at the Fermi level. Since our present observations already support the
$\mathbf{k}$-dependent hybridization, investigating the dispersion of the
\ket{\pm5/2} doublet along the $c$ axis would be desirable.

We note that the observation of the second
Fermi-momenta $k^{\beta}_{\mathrm{F2}}$ may also be consistent with the model
of deconfinement of the $f$ electrons~\cite{Senthil_PhysicaB2005}, whose
application to \bYAB~has been proposed by Hackl and
Thomale~\cite{Hackl_PRB2011}. 
In this model, $k^{\beta}_{\mathrm{F2}}$ hosts the critical fluctuations,
while carriers at $k^{\beta}_{\mathrm{F1}}$, labeled `cold',
exhibit a FL behavior. Naively, this is compatible with the similar observation
of the Fermi-momenta $k^{\alpha}_{\mathrm{F}}$ in the FL of \aYAB.
Nevertheless, we did not observe any direct sign of
such deconfinement in our data; from the ARPES point of view,
measurements at lower temperature are necessary
to follow the coherence of the \ket{\pm5/2} and to settle this question. 
In fact, previous electrical and thermal transport
measurements~\cite{Sutherland_PRB2015} discards this scenario by highlighting a
discrepancy with the expectation advanced by Hackl and Thomale of a
characteristic maximum in the Wiedemann-Franz ratio~\cite{Hackl_PRB2011}. 

Finally, we note that the band structure observed in the present ARPES
measurement is consistent with the theory of electron-spin resonance~(ESR)
in \bYAB~by Ramires {\it et al.}~\cite{Ramires_PRL2014}. 
Indeed, in addition to the unconventional QC, \bYAB~shows a singular ESR signal,
characterized by hyperfine satellites at low temperature, and by the constancy
of the ESR signal intensity~\cite{Holanda_PRL2011}. 
The work of Ramires {\it et al.} reproduces first the hyperfine satellites by
including Yb atoms with nonzero nuclear spin into the Kondo screening, and then
the constancy of the signal intensity by a crystal electric field comparable to
the Kondo hybridization strength~\cite{Ramires_PRL2014}. 
The band structure we observe in ARPES, with the hybridization of about 4~meV
of the same order as the crystal electric field  $\sim 9$~meV, directly
confirms the theoretical assumption in Ref.~\onlinecite{Ramires_PRL2014}. 
Additionally, the absence of tangible sign of the Kondo-like hybridization
within the experimental resolution of our ARPES measurements of \aYAB,
along with the absence of hyperfine satellites in its ESR
signal~\cite{Holanda_JoP2015}, tends to reinforce this model.  

\section{Conclusions}
In conclusion, we report an ARPES study of both $\beta$- and \aYAB, with 
the key observation of a Kondo-like hybridization between
the crystal-field split Yb $f$ levels and an electronlike conduction band in
\bYAB. 
The crystal-field splitting inferred from our ARPES data has a magnitude of
$\sim 9$~meV, in good agreement with previous
estimation~\cite{Nevidomskyy_PRL2009}.
Based on this agreement, we are able to use the
theoretical arguments to identify the low-lying heavy band at $-4$~meV
below $E_\mathrm{F}$ as the ground state $|J_z = \pm5/2\rangle$ doublet, and the
band at $-13$~meV as the first excited crystal-field state, dominated by
$|J_z = \pm3/2\rangle$ of the Yb \mbox{$J = 7/2$} multiplet.

The observed hybridization gives rise to a small electronlike QP pocket whose
size and effective mass are consistent with quantum oscillations measurements
\cite{OFarrell_PRL2009}, together with a heavier Fermi sheet as the hybridized
bands overlap in energy. 
Importantly, this QP exhibits a binding energy and an onset of coherence
which both conform with the earlier indications
of a reduced Kondo coherence temperature
$\Tcoh \approx 40$~K~\cite{OFarrell_PRL2012,Matsumoto_PRB2011}. 
Thus, as evidenced by the Hall effect measurements, it is indeed these Fermi
sheets that are responsible for both Kondo lattice and QC behaviors
below $T^*$~\cite{OFarrell_PRL2012}. 

Measurements of \aYAB~further support our conclusion. 
Indeed, 
in \aYAB, we similarly observe the electronlike conduction band and the flat
band at $-13$~meV below $E_\mathrm{F}$, which we also identify as the
\ket{\pm3/2} state. 
However, where \bYAB~shows two anti-crossings of Yb $4f$ levels with the
dispersive conduction band, the bands in the $\alpha$ phase do not show, within
our resolution, any apparent sign of hybridization. 
This dichotomy strongly suggests that this hybridization plays a crucial role in
the unconventional quantum criticality in \bYAB,
whereas \aYAB~is not critical at ambient conditions.

While our observations cannot sharply infer the microscopic mechanism behind the
non-Fermi-liquid behavior in \bYAB, they provide an important insight for
future experimental and theoretical investigations. 
As we already mentioned, further ARPES measurements at lower temperature,
and in wider regions of the Brillouin-zone, will have the potential to shed more
light on this intriguing quantum critical behavior. 

	\begin{acknowledgments}
We thank S. Burdin, S. Watanabe and D. Malterre for helpful discussions. 
We thank P. Zhang for sharing his procedure to perform 2D curvature. 
We thank D. Evtushinsky and E. Rienks for their precious support on the 1-cubed
end-station. We thank HZB for the allocation of synchrotron radiation
beamtime and thankfully acknowledge the financial support by HZB. 
This work was supported by CREST, Japan Science and Technology Agency,
Grants-in-Aid for Scientific Research (Grant No. 16H02209, No. 25707030,
and No. 26105002),
by Grants-in-Aid for Scientific Research on Innovative Areas ``J-Physics"
(Grant No. 15H05882 and No. 15H05883) and Program for Advancing Strategic
International Networks to Accelerate the Circulation of Talented Researchers
(Grant No. R2604) from the Japanese Society for the Promotion of Science.
A.H.N. was supported by the grant No. DMR-1350237 from the U.S. National Science
Foundation.
\end{acknowledgments}

\bibliographystyle{apsrev4-1}
%

\onecolumngrid
\begin{center}\begin{figure}
  \insertFig{16cm}{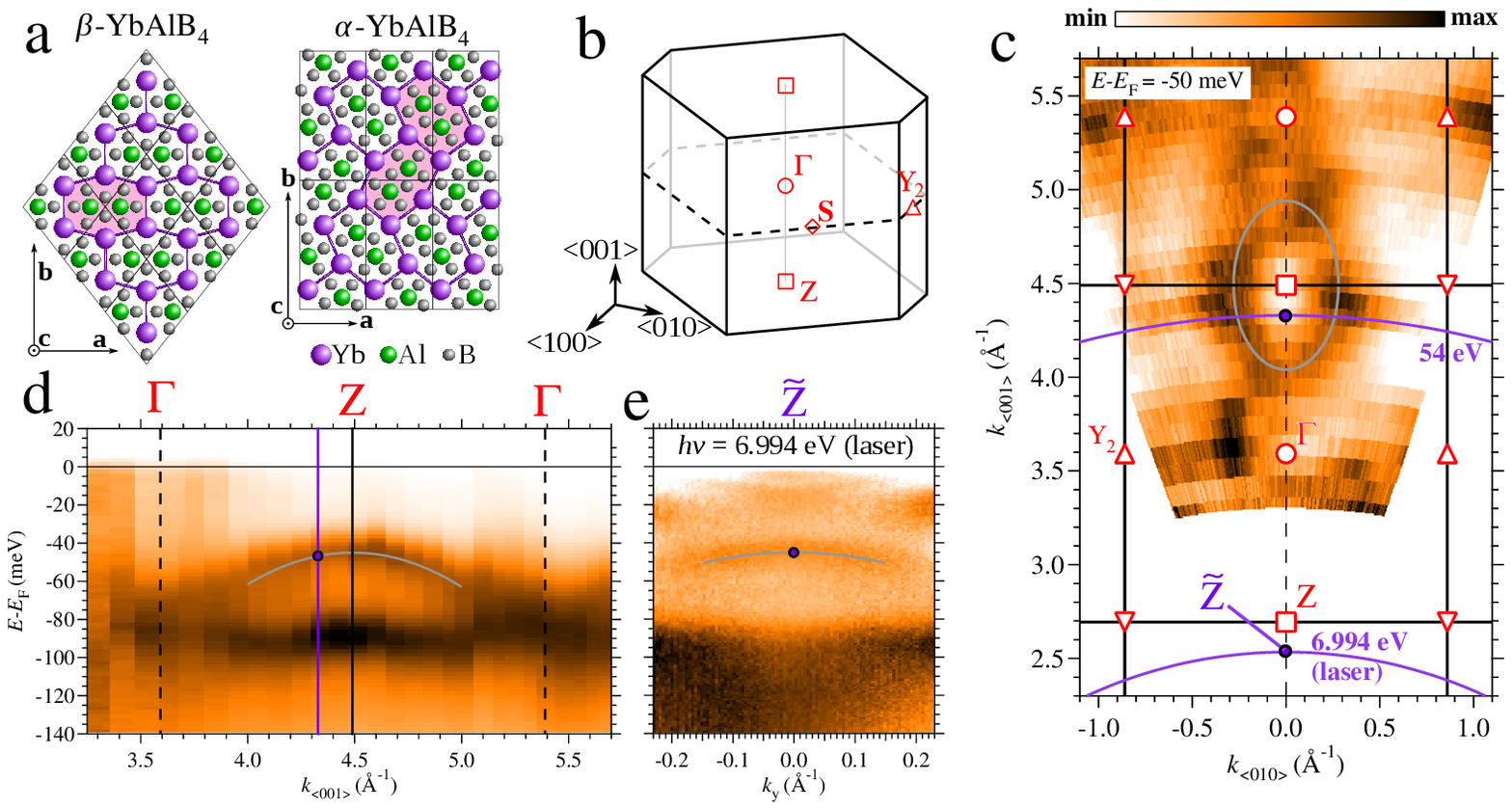}{kzDep}
          {
           a. Representation from the $c$ axis of the crystal structure of \bYAB
           (left) and \aYAB~(right). 
           b. Brillouin-zone of \bYAB. 
           c. Mapping in the (100)-plane, 50 meV below $E_\mathrm{F}$. Full
           black lines are the edges of the BZ. Purple circles show the position
           reached with a light of $h\nu = 6.994$~eV and $54$~eV. 
           d. Dispersion at normal emission, i.e., along the $\left<001\right>$
           direction. 
           e. Laser-ARPES spectra on \bYAB; $h\nu = 6.994$~eV. 
          }
\end{figure}\end{center}

\begin{center}\begin{figure}
  \insertFig{12.3cm}{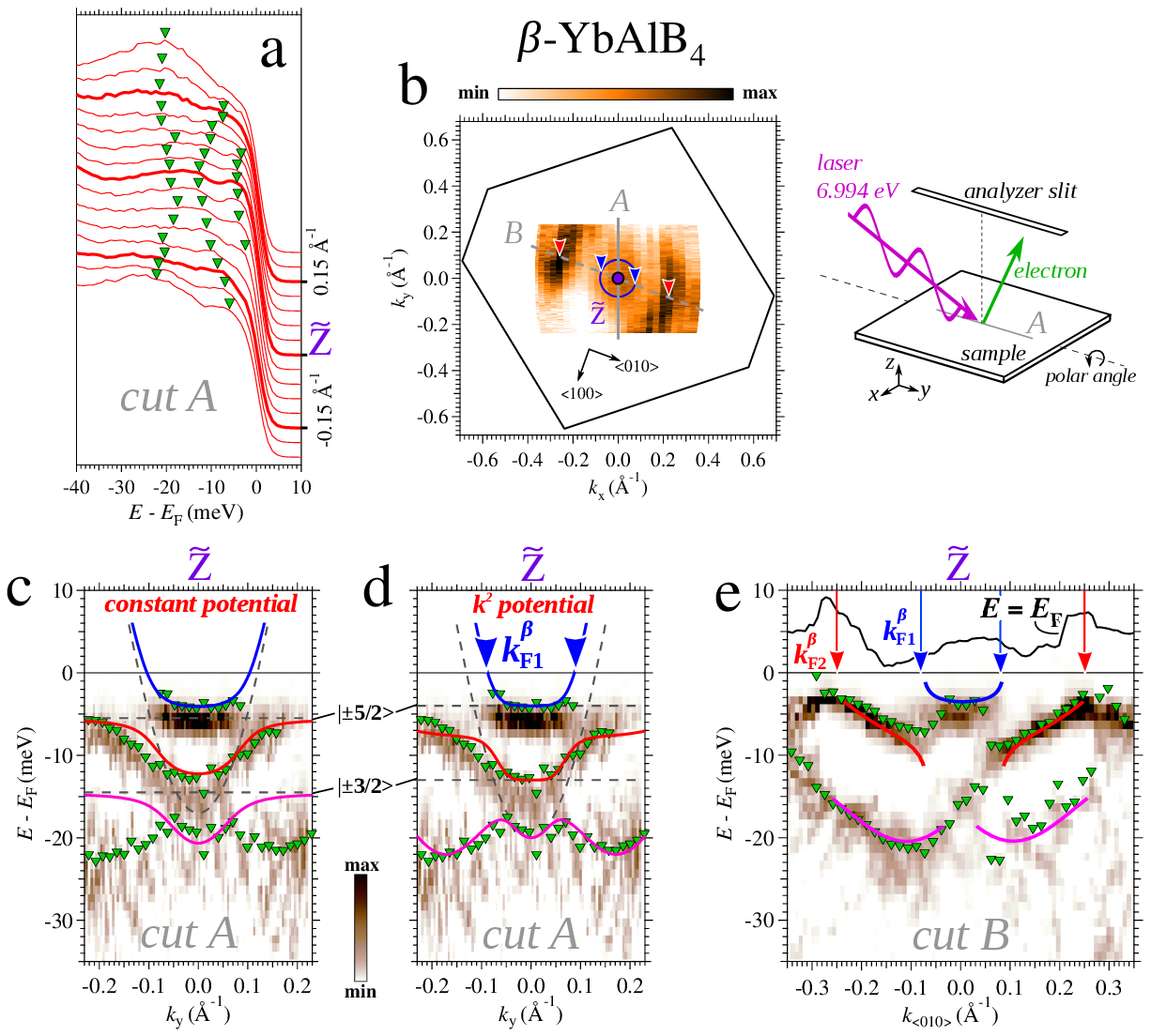}{disp5K}
          {
           Laser-ARPES on \bYAB. 
           a. Stack of EDCs next to $\Zt$, along the full gray line $A$ on
           panel~(b).
           Green markers are peak position from EDCs, fitted by Voigt with a
           quadratic background and cut by a Fermi-Dirac step. 
           Detailed fitting for
           $k_{\mathrm{y}} = -0.15,~0$ and $0.15$~\AA$^{-1}$ are illustrated
           with Fig. S2 of the Supplemental Material~\cite{SuppMat}.
           b. Fermi-surface mapping performed by tuning the polar angle.
           Sketch on the right illustrates the geometry of the measurements; 
           The analyzer slit was oriented along the line $A$. 
           The blue circle shows the little
           electronlike Fermi sheet, while the two red arrows
           show an outer Fermi sheet. Black lines are edges of the BZ. 
           c. Curvature of the ARPES spectra measured along
           the slit of the analyzer [full gray line on panel~(b)].
           Green markers are peak position as of panel~(a). 
           The blue, red and pink lines are the
           results of the modeling based on the constant hybridization of the
           original bands (represented by dashed black lines). 
           d. Same as (c) with a $\mathbf{k}$-dependent nodal potential. 
           See Supplemental Material for the raw data (Fig.~S3) and
           details of the hybridization models~\cite{SuppMat}. 
           e. Curvature of the cut along the $\left<010\right>$ direction
           [dashed gray line $B$ on panel~(b)].
           Green markers are peak position from EDC fit. Full blue, red and
           pink lines are guides to the eye. 
           See Fig. S3 of the Supplemental Material
           for the raw data~\cite{SuppMat}. 
          }
\end{figure}\end{center}

\begin{center}\begin{figure}
  \insertFig{6cm}{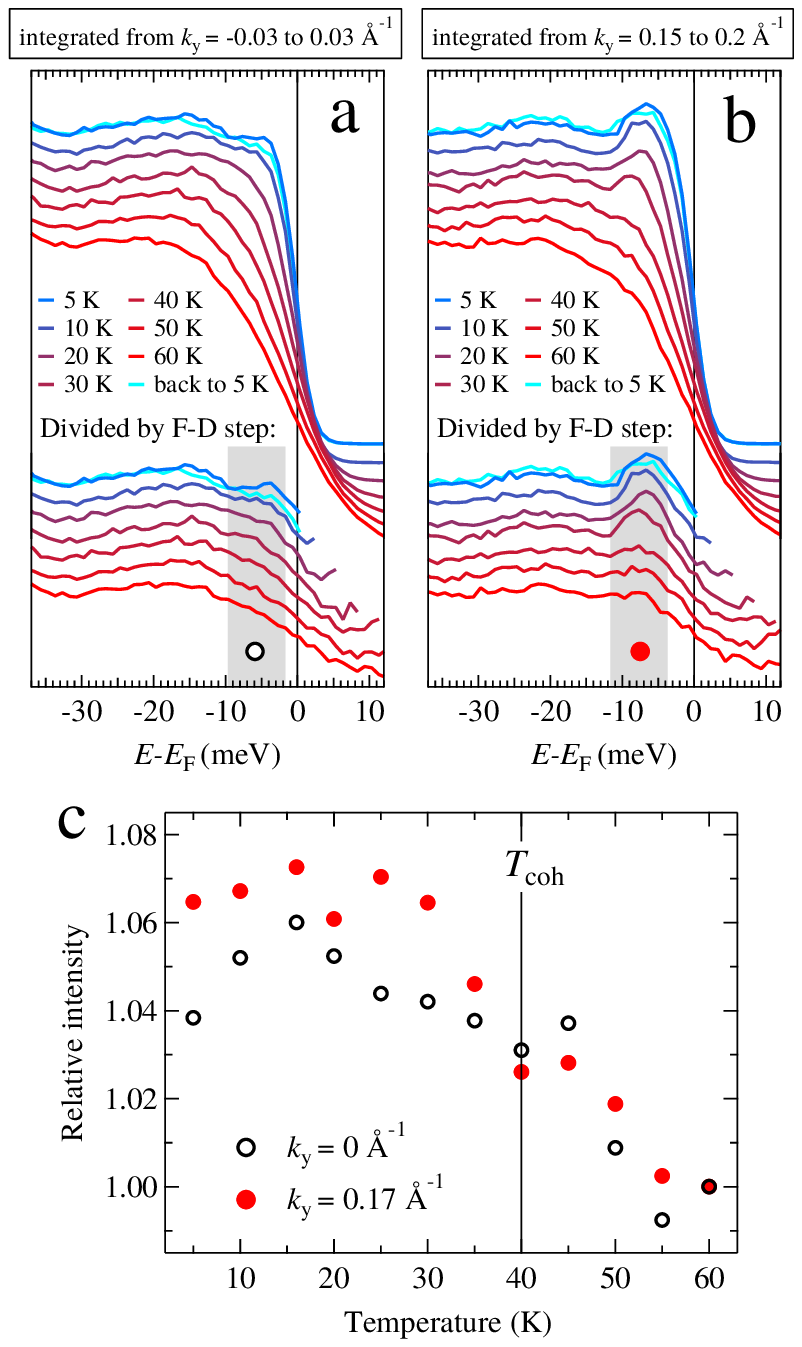}{Tdep}
          {
           Temperature dependence on \bYAB. 
           a. EDCs at $k_y = 0$\AA$^{-1}$ from $T=5$~K to $60$~K. 
           b. EDCs at $k_y = 0.17$\AA$^{-1}$ from $T=5$~K to $60$~K.
           [The peak's intensity difference with EDC in Fig. \ref{fig:disp5K}(a)
           is due to a slightly different orientation along the slit,
           as temperature dependence was performed on a different sample
            -- see Fig. S1 of the Supplemental Material
           for the corresponding color scale~\cite{SuppMat}.]
           c. Intensity, relative to the value at 60~K, of the inner (outer)
           peak from EDCs divided by the Fermi-Dirac step and integrated
           along the shade area of panel (a) [(b)].
          }
\end{figure}\end{center}

\begin{center}\begin{figure}
  \insertFig{7.75cm}{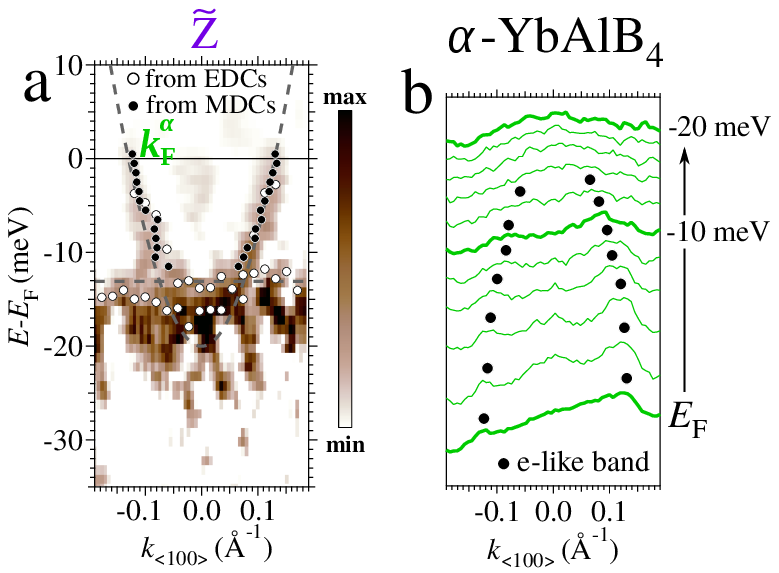}{aYAB}
          {
           Laser-ARPES on \aYAB. 
           a. Curvature of the spectra along the $\left<100\right>$ direction.
           Open and full circles are peak positions,
           extracted, respectively, from EDCs and MDCs. 
           See Fig. S3 of the Supplemental Material
           for the raw data~\cite{SuppMat}. 
           b. MDC stack of previous panel. Black markers are peak positions
           for the electronlike band. They were obtained by fitting MDCs by
           Voigts with a constant background. 
          }
\end{figure}\end{center}

\begin{center}\begin{figure}
  \insertFig{8cm}{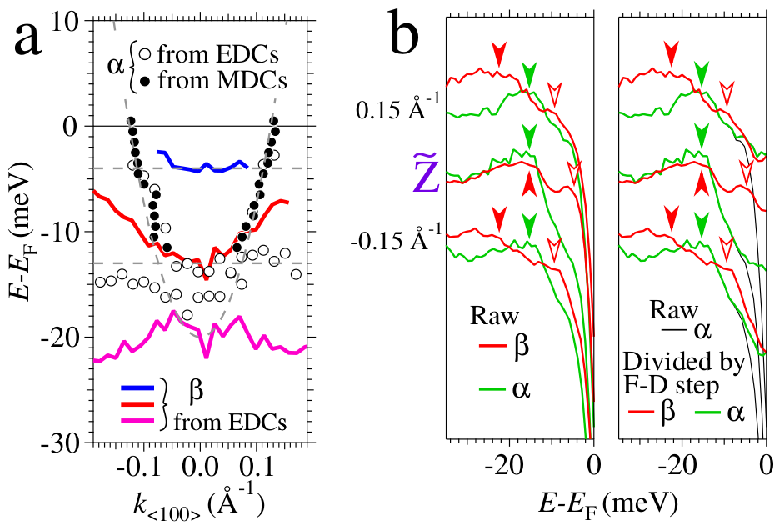}{bVSa}
          {
           a. MDC and EDC fits (open and full markers) from \aYAB~compared
           to EDC fit on \bYAB~(blue, red and pink lines). The broken black
           lines denote the unhybridized electronlike conduction band and Yb
           $4f$ flat bands from the same model as in Fig.~\ref{fig:disp5K}. 
           b. EDCs at $k = 0$;~$\pm0.15$\AA$^{-1}$ from \bYAB~(red) and
           \aYAB~(green). In the left panel is the raw, while the right panel
           shows the EDCs divided by the effective Fermi-Dirac step. Open arrows
           indicate the \ket{\pm5/2} level, while full arrows indicate the
           \ket{\pm3/2} level. 
          }
\end{figure}\end{center}


\onecolumngrid
\newpage
~
\newpage
~
\newpage
~
\newpage
\begin{center}
\textbf{\Large SUPPLEMENTAL MATERIAL}
\vspace{2cm}
\end{center}

\setcounter{figure}{0}
\renewcommand{\thefigure}{S\arabic{figure}}
\renewcommand{\theHfigure}{Supplement.\thefigure}
\textbf{\large Light polarization}

\begin{center}\begin{figure}[h]
  \insertFig{10cm}{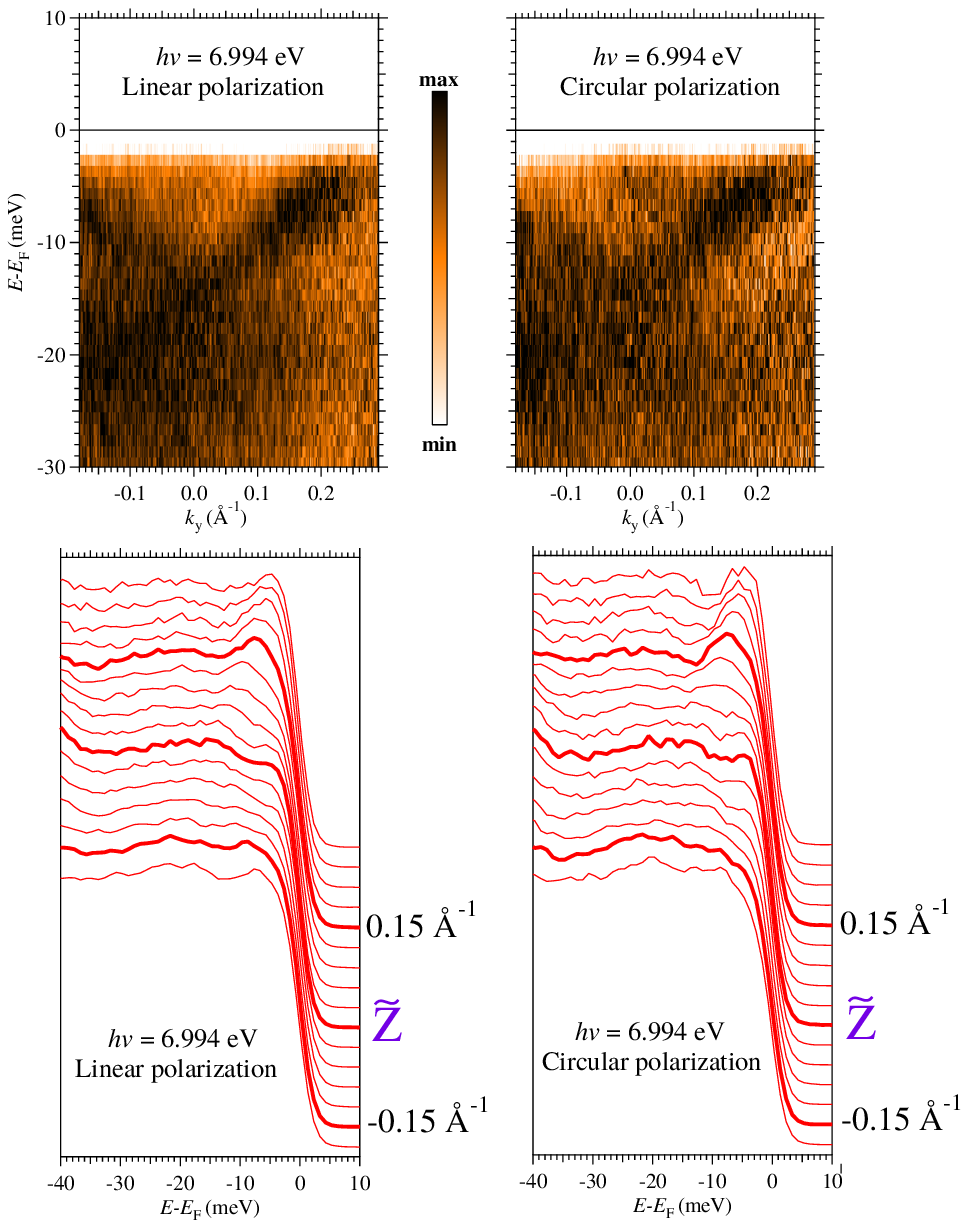}{pola}
          {
           Laser-ARPES spectra measured at normal emission, with linear
           polarized (left) and circular polarized (right) light. 
           For both polarization, the 3 same heavy bands can be observed as in
           the main text, with an anticrossing giving rise to the little 
           quasi-particle at about 4~meV. 
           Difference in the momentum distribution of the intensity compared to
           the main text is due to a slightly different orientation of the
           sample. 
          }
\end{figure}\end{center}

\newpage
\textbf{\large EDCs fits}

\begin{center}\begin{figure}[h]
  \insertFig{15.5cm}{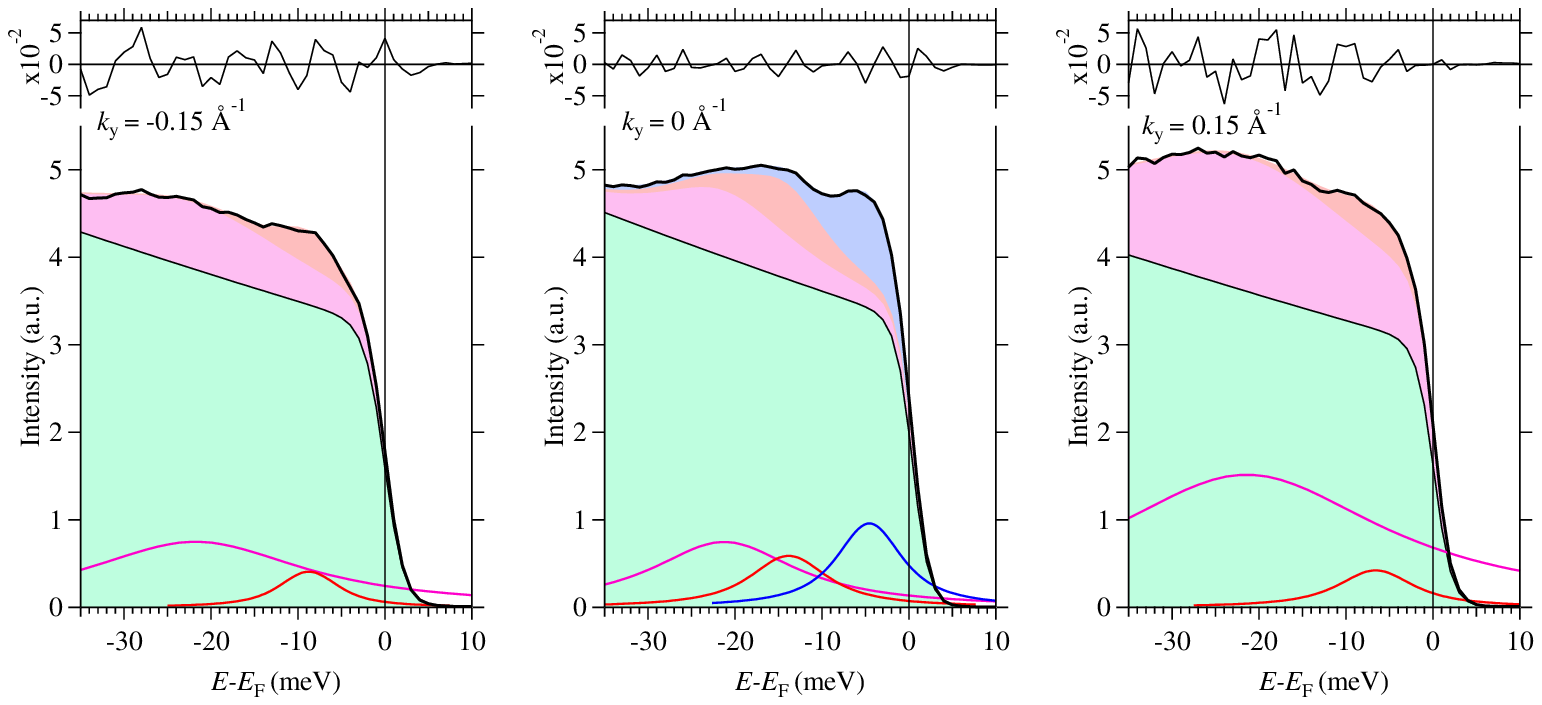}{voitFits}
          {
           Fit of EDCs from laser-ARPES on \bYAB~(see Fig.2 of the main text)
           at $k_\mathrm{y} = -0.15,~0$ and $0.15$~\AA$^{-1}$,
           from left to right. 
           The fit uses Voigt peaks with a quadratic background cut by
           a effective Fermi step. 
           Resulted Voigts are visible at the bottom of each graph, while their
           contribution to the fit represented with colored area. 
           Green areas depict contribution of the quadratic background. 
           Top curves are residual from the fitting. 
          }
\end{figure}\end{center}
\vspace{1cm}

\textbf{\large Raw Spectra}

\begin{center}\begin{figure}[h]
  \insertFig{12cm}{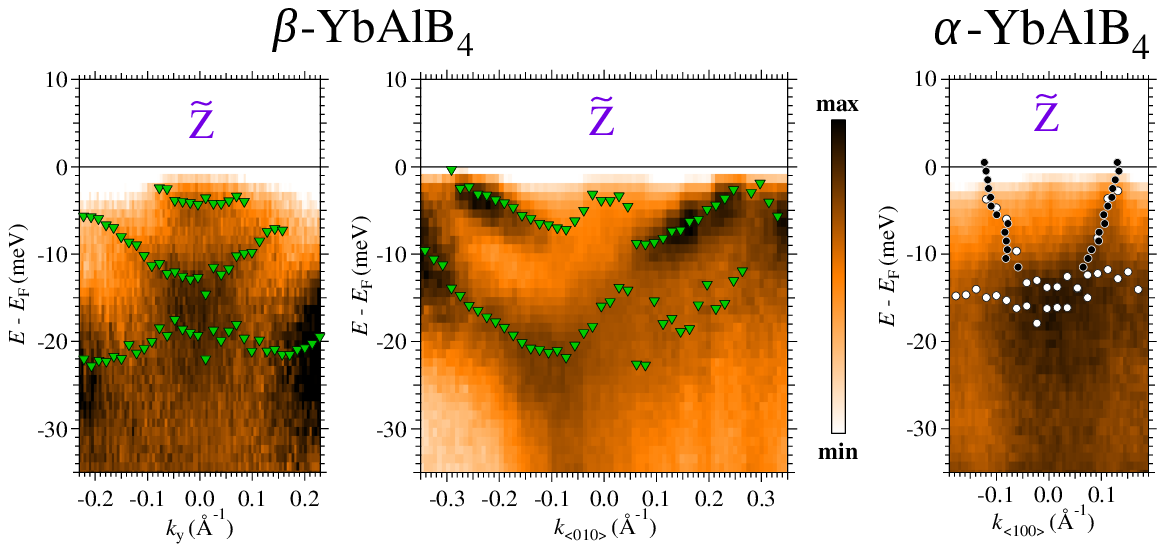}{rawData}
          {
           Raw spectra of the 2D curvature spectra shown in the main text. 
          }
\end{figure}\end{center}

\newpage
\textbf{\large Hybridization potential}\\

Models in the main text use a simple hybridization Hamiltonian, which can be
written as follow: 
$$
\mathcal{H}_{\mathrm{hyb}}=
  \left({ \begin{array}{ccc}
    \tilde{\epsilon}_{f,5/2} & 0                        & V_k \\
    0                        & \tilde{\epsilon}_{f,3/2} & V_k \\
    V_k^*                    & V_k^*                    & \epsilon_{c,k}
  \end{array} }\right)
$$
With $\tilde{\epsilon}_{f,5/2}$ and $\tilde{\epsilon}_{f,3/2}$($=\tilde{\epsilon}_{f,5/2}-9$~meV), the
renormalized energies of the Yb \ket{\pm5/2} and \ket{\pm3/2} $f$-levels respectively;
$\epsilon_{c,k} = \epsilon_{c0}+\frac{\hbar^2}{m_c}\frac{1}{a^2}
\left(1-\cos(ak_\mathrm{y})\right)$, 
the conduction band dispersion, with $a = 7.308$~\AA, the lattice parameter,
$m_c$ its relative mass, and $\epsilon_{c0}$ the bottom of the band;
and finally $V_k$, the hybridization potential. 

Parameters values are given in this table for both the constant and quadratic
hybridization toy models: 
$$
  \begin{array}{|c|c|c|}
    \hline
    ~                        & \mathrm{constant}  & \mathrm{\mathbf{k}-dependent} \\
    \hline
    m_c                      & 3 m_e              & 3 m_e \\
    \epsilon_{c0}            & -17~\mathrm{meV}   & -20~\mathrm{meV} \\
    \tilde{\epsilon}_{f,5/2} & -5.5~\mathrm{meV}  & -4~\mathrm{meV} \\
    \tilde{\epsilon}_{f,3/2} & -14.5~\mathrm{meV} & -13~\mathrm{meV} \\
    V_k                      & 4~\mathrm{meV}     & C\sin^2(ak_\mathrm{y}) \\
    \hline
  \end{array}    
$$
with $m_e$ the free electron mass, and $C = 16$~meV. 


\end{document}